\documentclass[review]{elsarticle}

\usepackage{lineno,hyperref}
\modulolinenumbers[5]

\usepackage{geometry}
\newgeometry{left=2cm,right=2cm,top=2.5cm,bottom=2.5cm}
\usepackage{amsmath}
\usepackage{amssymb}
\usepackage{amsbsy}
\usepackage{amsfonts}
\usepackage{amsopn}
\usepackage{amstext}
\usepackage{graphicx}
\usepackage{amssymb}
\usepackage{amsfonts}
\usepackage{amsmath}
\usepackage{graphicx}
\usepackage[english]{babel}
\usepackage{color}
\usepackage{slashed}
\usepackage{esint}
\usepackage[dvips]{epsfig}
\usepackage[dvips]{graphicx}
\usepackage{float}
\usepackage{units}
\usepackage{textcomp}
\usepackage{caption}
\usepackage{multirow}
\usepackage{mathrsfs}
\usepackage{bm}
\usepackage{amsmath}
\usepackage{amssymb}
\usepackage{amsthm}
\usepackage{amsfonts}
\usepackage{mathtools}
\usepackage{color}
\usepackage{graphicx}
\usepackage{dcolumn}
\usepackage{tabularx}
\usepackage{array}
\usepackage{booktabs}
\usepackage{subcaption}
\usepackage{graphicx} 
\usepackage{hyperref}
\usepackage{bbold}
\usepackage{systeme}
\usepackage{epstopdf}
\usepackage{hyperref}
\usepackage{slashed}
\usepackage{hyperref}
\usepackage{graphicx}
\usepackage{amssymb}
\usepackage{amsmath}
\usepackage{color}
\usepackage{adjustbox}

\usepackage[normalem]{ulem}

\newcommand{\be}{\begin{equation}}
\newcommand{\ee}{\end{equation}}
\newcommand{\ben}{\begin{eqnarray}}
\newcommand{\een}{\end{eqnarray}}

\journal{Annals of Physics}

\begin{document}

\begin{frontmatter}

\title{Quantum information entropy of heavy mesons in the presence of a point-like defect}

\author[inst1]{C. A. S. Almeida\footnote{Email address: carlos@fisica.ufc.br}}

\author[inst2,inst3,inst4]{C. O. Edet\footnote{Email address: collinsokonedet@gmail.com}}

\author[inst1]{F. C. E. Lima\footnote{Email address: cleiton.estevao@fisica.ufc.br}}

\author[inst3,inst5]{N. Ali\footnote{Email address: norshamsuri@unimap.edu.my}}

\author[inst6]{and M. Asjad\footnote{Email address: muhammad.asjad@ku.ac.ae}}

\address[inst1]{Universidade Federal do Cear\'{a} (UFC), Departamento de F\'{i}sica, Campus do Pici, Fortaleza-CE, 60455-760, Brazil.}

\address[inst2]{Institute of Engineering Mathematics, Universiti Malaysia Perlis, 02600 Arau, Perlis, Malaysia.}

\address[inst3]{Faculty of Electronic Engineering Technology, Universiti Malaysia Perlis, Malaysia.}

\address[inst4]{Department of Physics, Cross River University of Technology, Calabar, Nigeria.}

\address[inst5]{Advanced Communication Engineering (ACE) Centre of Excellence, Universiti Malaysia Perlis, 01000 Kangar, Perlis, Malaysia.}

\address[inst6]{Department of Mathematics, Khalifa University, Abu Dhabi 127788, United Arab Emirates.}

\begin{abstract}
Using Schr\"{o}dinger's formalism, we investigate the quantum eigenstates of the heavy mesons trapped by a point-like defect and by Cornell's potential. One implements this defect to the model considering a spherical metric profile coupled to it. Furthermore, the Nikiforov-Uvarov method is applied to theory to study the quantum eigenstates of the heavy mesons. To calculate the quantum information entropy (QIE), one considers the wave functions that describe the charmonium and bottomonium states. To explore the QIE, we use the well-known Shannon's entropy formulated at the position and reciprocal space. The analysis of the QIE gives us relevant information about how the quantum information change with the variation of the point-like defect. Consequently, considering the Bialynicki-Birula and Mycielski (BBM) relation, we show how this defect influences the quarkonium position and momentum uncertainty measures.

\end{abstract}
\end{frontmatter}

\section{Introduction}
The growing interest in non-relativistic quantum mechanical systems is notorious \cite{SDong1,SDong2,GSun1,GSun2}. This interest is because the results from quantum mechanics give us a good prediction of some phenomenological data \cite{Quigg,He,Band}. Furthermore, these systems are the first step toward understanding more complex models \cite{Bhat}. Thus, considering non-relativistic quantum mechanical systems, we found several studies in the literature \cite{COEdet1,COEdet2}. For example, one can find some studies on theoretical measurements of mass spectra of heavy mesons \cite{Omugbe}, harmonic oscillators \cite{Dekker,Chang,Jafarov}, potential wells \cite{FCLima1,Ruostekoski}, and solid-state physics problems (e. g., position-dependent mass problems) \cite{Dutra1,Dutra2,Renan}.

Generally, depending on the type of interaction used to describe the quantum-mechanical system, there may be some difficulties in the analytical description. Thus, to bypass these difficulties, some techniques are applied. Among these techniques, there is the well-known Nikiforov-Uvarov (NU) method \cite{GSun1,Willian,Omugbe2,Miranga}, the Nikiforov-Uvarov Functional Analysis (NUFA) \cite{Inyang,Ikot,Okon}, the Laplace transformation method (LTM) \cite{Shady}, the asymptotic interaction method \cite{COEdet3}, the WKB approximation method \cite{EOmugbe3}. However, it is relevant to highlight that among the techniques mentioned, the NU method stands out \cite{karayer,Gonul,NU}. Indeed, the NU method stands out because it allows parameterizing Schr\"{o}dinger's equation \cite{Tezcan}, making it possible to find the solution for different physical potentials of interest \cite{Tezcan}. Thus, we will conveniently use the NU method in this work.

A relevant class of particles is mesons. The historical framework of these particles began in 1974 \cite{Aubert}. It was in 1974 that the Brookhaven National Laboratory announced the discovery of a new particle called $J$ \cite{Aubert}. Simultaneously, the Stanford Linear Accelerator announced the existence of another particle called $\psi$ \cite{Glashow}. A relevant feature of these particles found is that they have similar properties. So has become possible to interpret these particles as a new quark (namely, a charm quark). With the discovery of these particles, it was possible to explain the suppression of weak kaon decays that change the flavor \cite{Herb,Kelly}. Posteriorly, the fourth quark existence, the so called bottom quark, was confirmed \cite{Herb,Kelly}. Only in 1978 Cornell's model was developed. In its original proposal, Cornell's potential took into account the quark-antiquark interaction in the heavy sector \cite{Eichten1,Eichten2,Eichten3,Eichten4}. Essentially the initial assumptions were that interactions arise $SU(3)$ color gauge symmetry with flavor broken only by quark masses. Besides, this type of interaction admitted contributions from the Coulomb interaction (induced by a gluon exchange interaction). Furthermore, Cornell's potential considered contributions from a phenomenological interaction of confinement, considered linear. The interactions were independent of flavor and spin, thus implementing the well-known Heavy Flavor Symmetry and Heavy Quark Spin Symmetry \cite{Eichten1,Eichten2,Eichten3,Eichten4}.

Seeking to investigate the quarkonia eigenstates, we will consider Cornell's potential in our theory. This potential proves to be effective in explaining quark confinement. Furthermore, this potential explains the mass of the quarkonium states and its relation with the angular momentum of the hadron (known as the Regge trajectory) \cite{Eichten1,Eichten2,Eichten3,Eichten4}. Briefly, the Cornell potential has the following form:
\begin{align}\label{cornell}
    V(r)=-\frac{4}{3}\frac{\alpha_s}{r}+\sigma r+v_0,
\end{align}
where $r$ is the effective radius of the quarkium state, $\alpha_s$ is the running coupling, $\sigma$ is the QCD string tension, and $v_0\simeq -0.3$GeV. Note that this potential has two contributions. The first contribution (i.e., $-\frac{4}{3}\frac{\alpha_s}{r}$) dominates at short distances, in general for $r<0.1$fm \cite{Brambilla}. This short-range contribution is known as Yukawa's potential. In this case, this contribution is due to the exchange of a gluon between the quark and its antiquark. Meanwhile, the second potential contribution (i.e., $\sigma r$) is the linear confinement term responsible for describing the non-perturbative QCD effects that result in color confinement. In this scenario, $\sigma$ is the tension of the QCD string that forms when the gluonic field lines collapse in a flow tube. One estimates that the tension value of the QCD string is $\sigma\simeq 0.18$GeV$^{2}$ \cite{Deur}. Furthermore, it is important to mention that Cornell's potential (\ref{cornell}) has been gaining supporters in recent years. For example, the Cornell potential has been used in spectroscopy studies of mesons \cite{Soni} and S-wave heavy quarkonia \cite{Chung}, among other studies.

Not far from this discussion, we found, in literature, some works discussing the influence of defects in quantum eigenstates \cite{COEdet4,Ali,Ahmed}. Indeed, the study of topological defects in quantum mechanics problems has aroused the interest of several researchers \cite{Silva,Maia,Bakke}, and this interest is due to its wide application. Generally speaking, there are three classes of defects, i. e., the cosmic string defect, the global monopole, and domain walls \cite{Kibble,Vilenkin}. Usually, one-dimensional cosmic string defects \cite{Zare} and global monopoles \cite{Barriola,Bennett,Ahmed2} apply widely to quantum mechanical systems. Indeed, in this scenario, these defects were implemented, for example, in studies of harmonic oscillators \cite{Furtado,Vitoria}, charged particle scattering \cite{Oliveira}, and models with exotic interaction \cite{Nwabuzor}. Considering these applications arises the question: How does the monopole-like defect influence the eigenstate of heavy mesons? Throughout this manuscript, we will seek to answer this question.

Quantum information entropy (QIE) is a useful tool in studies on quantum information and uncertainty measurements in quantum-mechanical systems. Indeed, the QIE arises from a reinterpretation of Shannon's information entropy \cite{Shannon}. Essentially, this reinterpretation, i. e., the QIE, and the Bialynicki-Birula and Mycielski \cite{BBM} relation, gives us a generalized uncertainty measure of Heisenberg's principle. Due to this, the QIE has gained searchlight in studies on quantum-mechanical systems. Thus, one can find several works discussing the information of quantum-mechanical systems through the QIE formalism. For example, some applications of the QIE in the quantum system appear in investigations of hyperbolic potential wells \cite{Gil}, the Aharonov-Bohm effect \cite{COEdet5}, and effective mass problems \cite{FCLima1}.

Considering the applications presented, two questions naturally arise. First, how do point-like global monopole (PGM) defects change the mesons' eigenstates? Second, how does the quantum information modifies due to this defect? As far as we know, our work is the first to address these questions. Thus, the main purpose of this paper is to study the QIE of heavy mesons in the presence of a point-like global monopole defect.

To reach our purpose, we organize the work as follows: In Sec. II, the quantum description of heavy mesons is exposed, assuming Cornell's potential. In Sec. III, we present a discussion and numerical results of the QIE of the Charmonium and Bottomonium states of heavy mesons. Finally, in Sec. IV, the findings are announced.

\section{Quantum description of heavy mesons}
\label{II}

As discussed in Ref. \cite{Barriola}, allow us to start by considering background spacetime with PGM defect. Particularly, let us assume a spherically symmetric spacetime described by the line element
\begin{align}\label{metric}
    ds^2=-dt^2+\eta_{ij}dx^idx^j, \hspace{0.5cm} \text{with} \hspace{0.5cm} i,j=1,2,3,
\end{align}
where the metric tensor $\eta_{ij}$ is
\begin{align}
    \eta_{ij}=\begin{pmatrix}
    \frac{1}{\alpha^2} & 0 &0\\
    0 & r^2 & 0\\
    0 & 0 & r^2\sin^2\theta
    \end{pmatrix} \hspace{0.5cm} \text{and} \hspace{0.5cm}
    \eta^{ij}=\begin{pmatrix}
    \alpha^2 & 0 &0\\
    0 & \frac{1}{r^2} & 0\\
    0 & 0 & \frac{1}{r^2\sin^2\theta}
    \end{pmatrix}.
\end{align}
Here the $\alpha$ parameter is related to the PGM defect, which depends on the energy scale. Mathematically, $\alpha^2=1-8\pi G\lambda^2$ where $\lambda$ is the energy scale and $G$ the gravitational constant.

Indeed, several researchers have adopted the PGM defect in their studies. For example, in Ref. \cite{AOliveira}, to investigate spin-0 particles in the presence of dyons and Aharonov-Bohm effect with scalar interaction, the authors adopted the existence of a PGM defect. Furthermore, Ahmed \cite{Ahmed} discussed the topological effects produced by this background on spin-0 particles subject to the Kratzed potential. Particularly the interest in this defect is because it induces topological effects that influence the particle's quantum dynamics. Motivated by this, we consider that spinless mesons are in the spacetime (\ref{metric}).

Considering the non-relativistic scenario, one writes Schr\"{o}dinger's equation as
\begin{equation}\label{eq4}
-\frac{\hbar^2}{2 \mu} \nabla^2_{LB}\psi(\textbf{r},t)+V(r,t)\psi(\textbf{r},t)=i \hbar \frac{\partial{\psi(\textbf{r},t)}}{\partial{t}}.
\end{equation}
Here, $\mu$ is the particle's mass and $\nabla^2_{\text{LB}}$ is the Laplace-Beltrami operator defined as
\begin{align}\label{Laplace}
    \nabla^2_{LB}\equiv\frac{1}{\sqrt{g}}\partial_{i}(\sqrt{g}g^{ij}\partial_{j}),
\end{align}
where $g\equiv$ det$(g_{ij})$. Furthermore, we will assume, in principle, that the interaction $V(\textbf{r},t)=V(r)$, i. e., an arbitrary central potential.

Assuming the Laplace-Beltrami operator (\ref{Laplace}) and the metric signature (\ref{metric}), we write Schr\"{o}dinger's equation as follows:
\begin{equation}\label{CSchrodinger}\nonumber
-\frac{\hbar^2 }{2 \mu r^2} \left[ \alpha^{2} \frac{\partial
}{\partial {r}}\left(r^{2}\frac{\partial
}{\partial {r}}\right) +\frac{1}{\sin{\theta}} \frac{\partial}{\partial {\theta}}\left(\sin{\theta}\frac{\partial}{\partial {\theta}}\right)+ \frac{1}{\sin^2{\theta}}\left(\frac{\partial^2}{\partial{\varphi^2}}\right) \right]\psi(r, \theta,\varphi,t)
+V (r) \psi(r, \theta,\varphi,t) = i \hbar \frac{\partial{\psi(r, \theta,\varphi,t)}}{\partial{t}}.
\end{equation}

Allow us to particularize our study for the case of spinless heavy mesons. In this case, the interaction of the theory is
\begin{align}\label{Potential}
    V(r)=W_1 r-\frac{W_2}{r}.
\end{align}
This interaction is called Cornell's potential. Essentially, Cornell's potential is employed to model the quarkonium interaction. Besides, one finds, in literature, several investigations on quantum-mechanics systems adopting Cornell's potential, e. g., see Refs. \cite{Kumar,ANIkot}.

To study the wave eigenfunctions that the theory describes, let us assume that the particular solutions of Eq. (\ref{CSchrodinger}) in terms of the eigenvalues of the angular momentum operator $\hat{L}^2$ are
\begin{equation}\label{rpsi}
\psi(\textbf{r}, t)=\text{e}^{-\frac{iE_{nl}t}{\hbar}} \frac{R_{nl}(r)}{r} Y_{l m}(\theta,\varphi)
\end{equation}
where $Y_{l m}(\theta,\varphi)$ are spherical harmonics and $R_{nl}(r)$ is the radial wave eigenfunction.

Substituting Eq. (\ref{rpsi}) into (\ref{CSchrodinger}), we have that the radial part of Schr\"{o}dinger's equation for Cornell's potential in the presence of the PGM defect is
\begin{align}\label{REq}
\dfrac{d^{2}R_{n l}(r)}{dr^2}+\left[\frac{2 \mu E_{n l}}{\alpha^2 \hbar^2}-\frac{2 \mu W_{1} r}{\alpha^2 \hbar^2}+\frac{2 \mu W_2}{\alpha^2 \hbar^2 r}-\frac{l (l+1)}{\alpha^2 r^2}\right]R_{n l}(r)=0.
\end{align}

Eq. (\ref{REq}) is not solvable in its current form. To solve this, allow us to adopt $r\to x^{-1}$. This change of coordinates leads us to
\begin{equation}
\dfrac{d^{2}R_{n l}(x)}{dx^2}+\frac{2}{x}\dfrac{dR_{n l}(x)}{dx}+\frac{1}{x^4}\left[\frac{2 \mu E_{n l}}{\alpha^2 \hbar^2}-\frac{2 \mu W_{1}}{\alpha^2 \hbar^{2}x}+\frac{2 \mu W_{2}x}{\alpha^2 \hbar^2}-\frac{l (l+1) x^2}{\alpha^2}\right]R_{n l}(x)=0.
\end{equation}

Here let us implement an approximation scheme (AS) on the term $\frac{W_1}{x}$ by assuming that there is a characteristic radius $r_0$ of the meson. In this case, one obtains the expansion of $\frac{W_1}{x}$  in a power series around $r_0$, i. e., around $\delta \equiv 1/r_0$, up to the second order. By setting $y=x-\delta$ around $y=0$, we obtain that
\begin{align}\label{NUEq}
\dfrac{d^{2} R_{n l} (x)}{dx^2}+\frac{2 x}{x^2}\dfrac{d R_{n l} (x)}{dx}+\frac{1}{x^4} \left[-\tilde{\epsilon}+\tilde{\beta}_{1}x-\tilde{\beta}_{2}x^2\right]R_{nl}(x)=0,
\end{align}
where
\begin{align}
    -\tilde{\epsilon}=\frac{2 \mu E_{n l}}{\hbar^{2} \alpha^2} - \frac{6 \mu W_{1}}{\alpha^{2} \hbar^{2} \delta},\quad \tilde{\beta}_{1}=\frac{2 \mu W_2}{\alpha^2 \hbar^2}+\frac{6 \mu W_1 }{\alpha^2 \hbar^2 \delta^2}, \quad \text{and} \quad \tilde{\beta}_{2}=\frac{2 \mu W_1}{\alpha^{2} \hbar^{2} \delta^{3}}+\frac{l(l+1)}{\alpha^2}.
\end{align}
Eq. (\ref{NUEq}) is the solvable form of the NU method. The major equation closely related to this method is
\begin{equation}\label{eq9}
P^{\prime \prime}(x)+\frac{\widetilde{\tau}(x)}{\sigma(x)}P^{\prime}(x)+\frac{\widetilde{\sigma}(x)}{(\sigma(x))^2}P(x)=0.
\end{equation}

\begin{figure*}[!ht]
    \begin{center}
    \includegraphics[height=4.5cm,width=5cm]{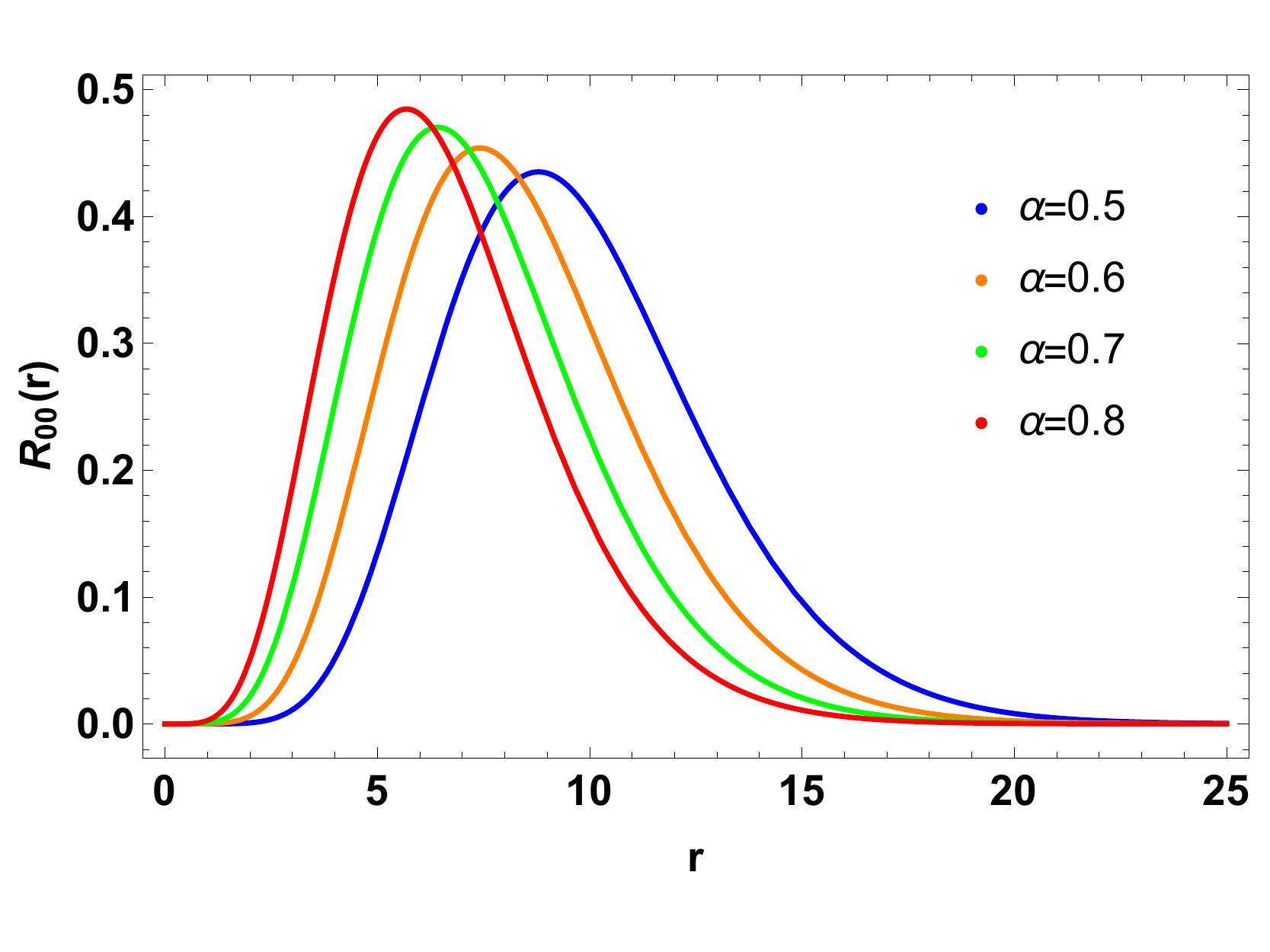}
    \includegraphics[height=4.5cm,width=5cm]{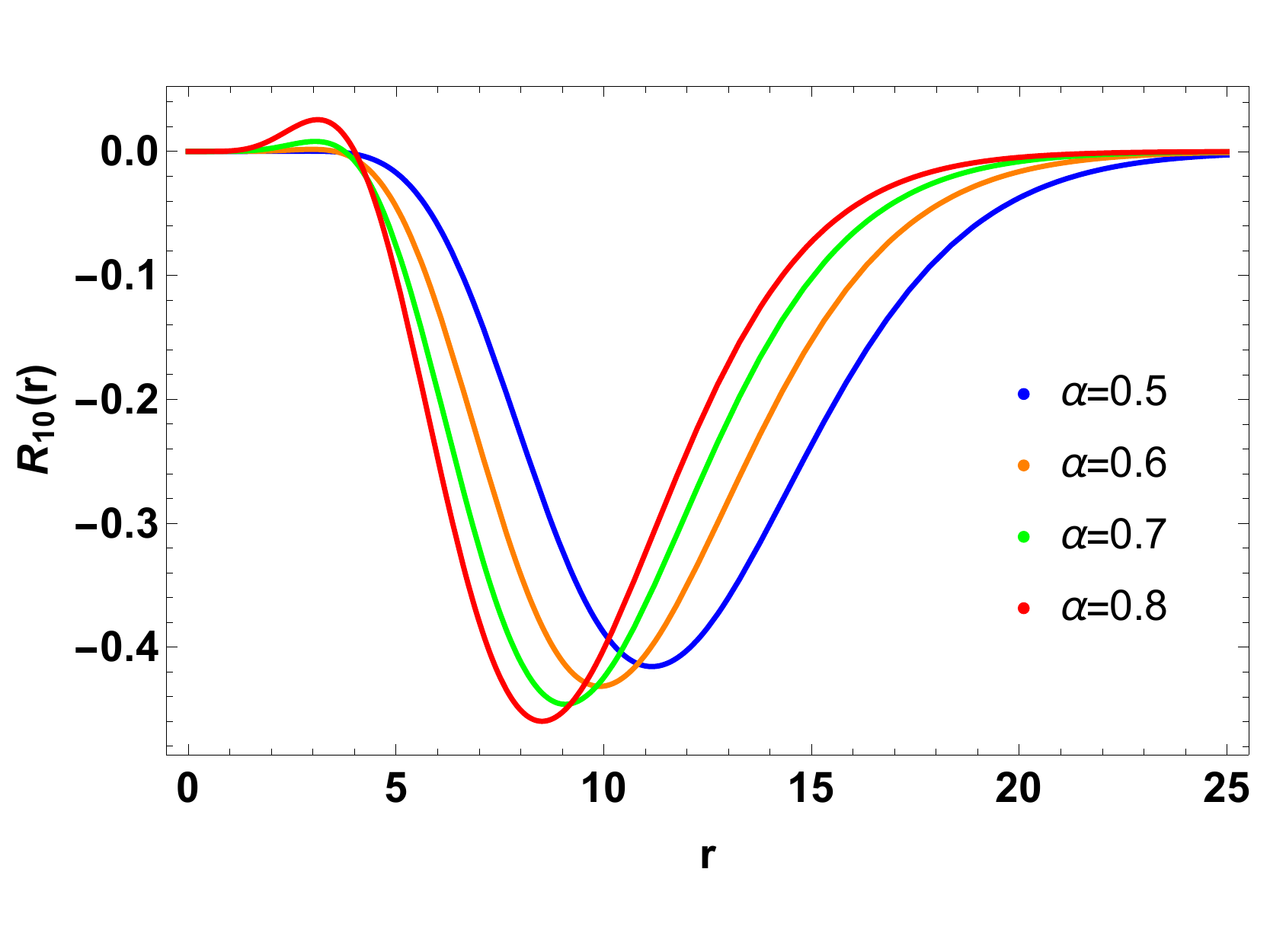}
    \includegraphics[height=4.5cm,width=5cm]{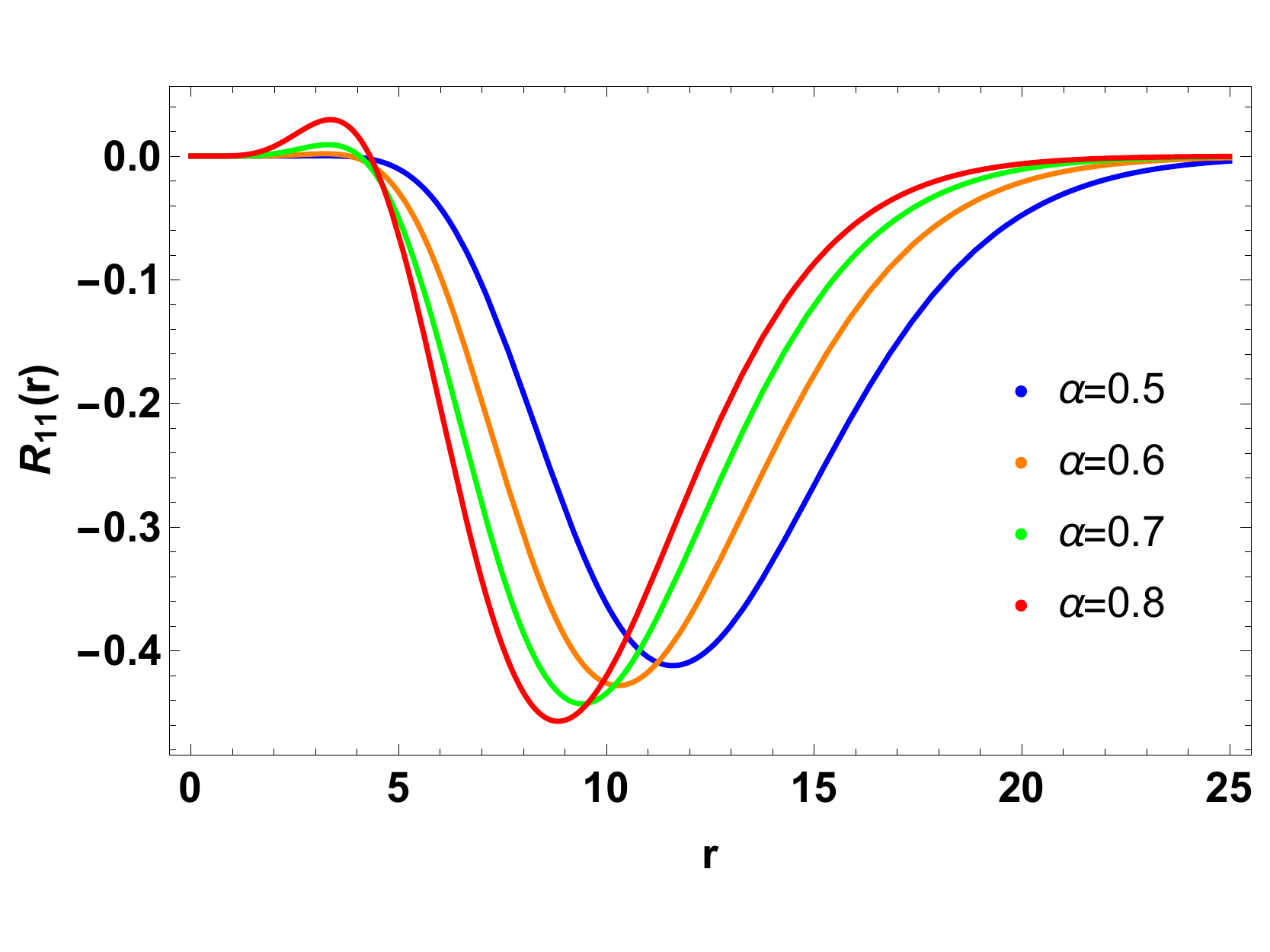}\\ 
\vspace{-0.3cm}
    (a) \hspace{4.5cm} (b) \hspace{4.5cm} (c)
    \end{center}\vspace{-0.5cm}
    \caption{Radial wave eigenfunctions of heavy mesons (Charmonium) when $m_{c}=1.209\,\text{Gev}, \; \mu= 0.6045\,\text{Gev}, \; W_{1}=0.20\,\text{Gev}, \; W_{2}=1.244\,\text{Gev}, \;\delta=0.231\,\text{Gev}, \text{and}\; \hbar=1$. Fig. (a) corresponds to the state $n=0$ and $l=0$. On the other hand, Fig (b) is the state $n=1$ and $l=0$. The state $n=1$ and $l=1$ is in Fig. (c).}
    \label{fig1}
\end{figure*}

\begin{figure*}[!ht]
    \begin{center}
    \includegraphics[height=4.5cm,width=5cm]{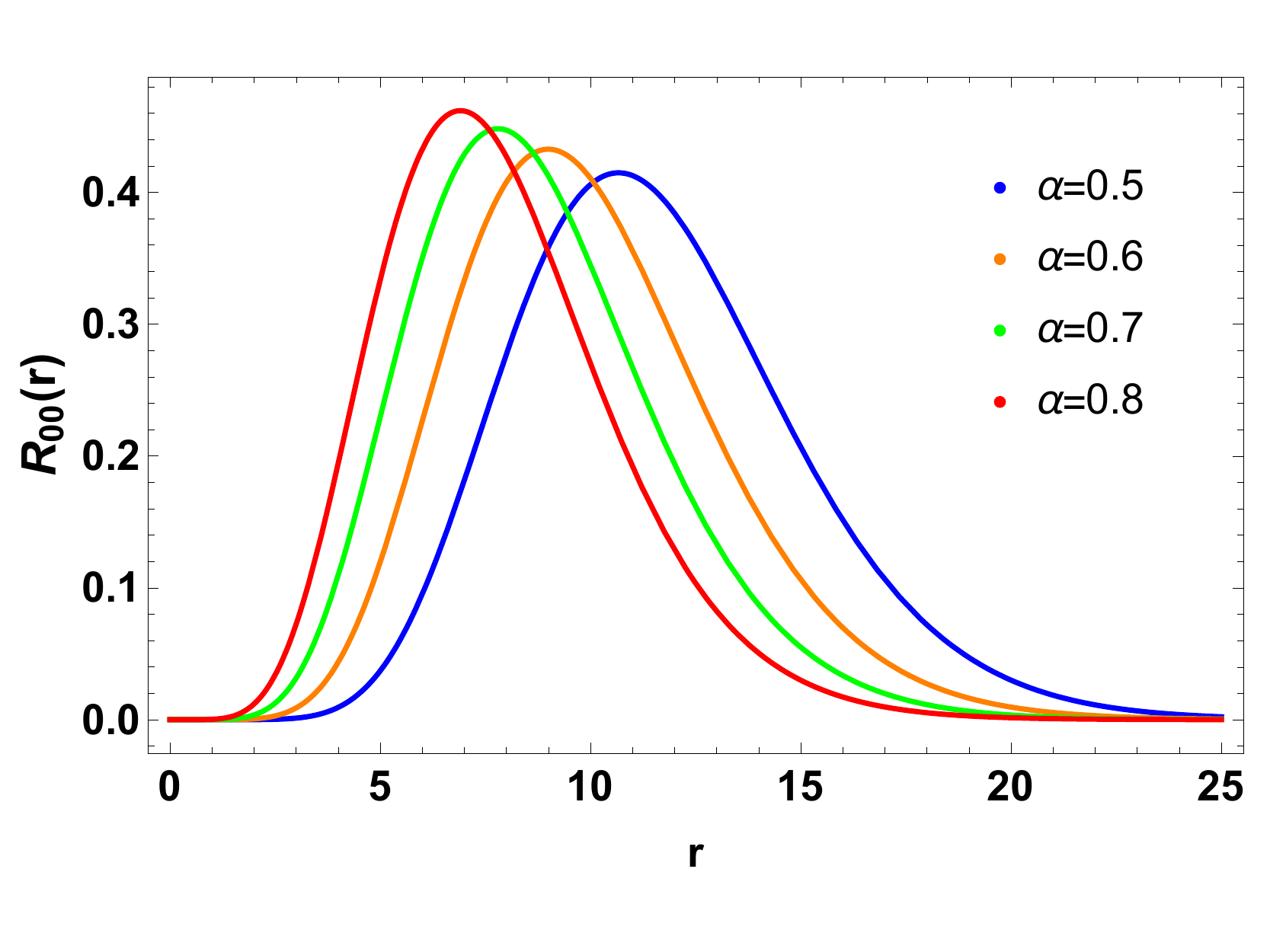}
    \includegraphics[height=4.5cm,width=5cm]{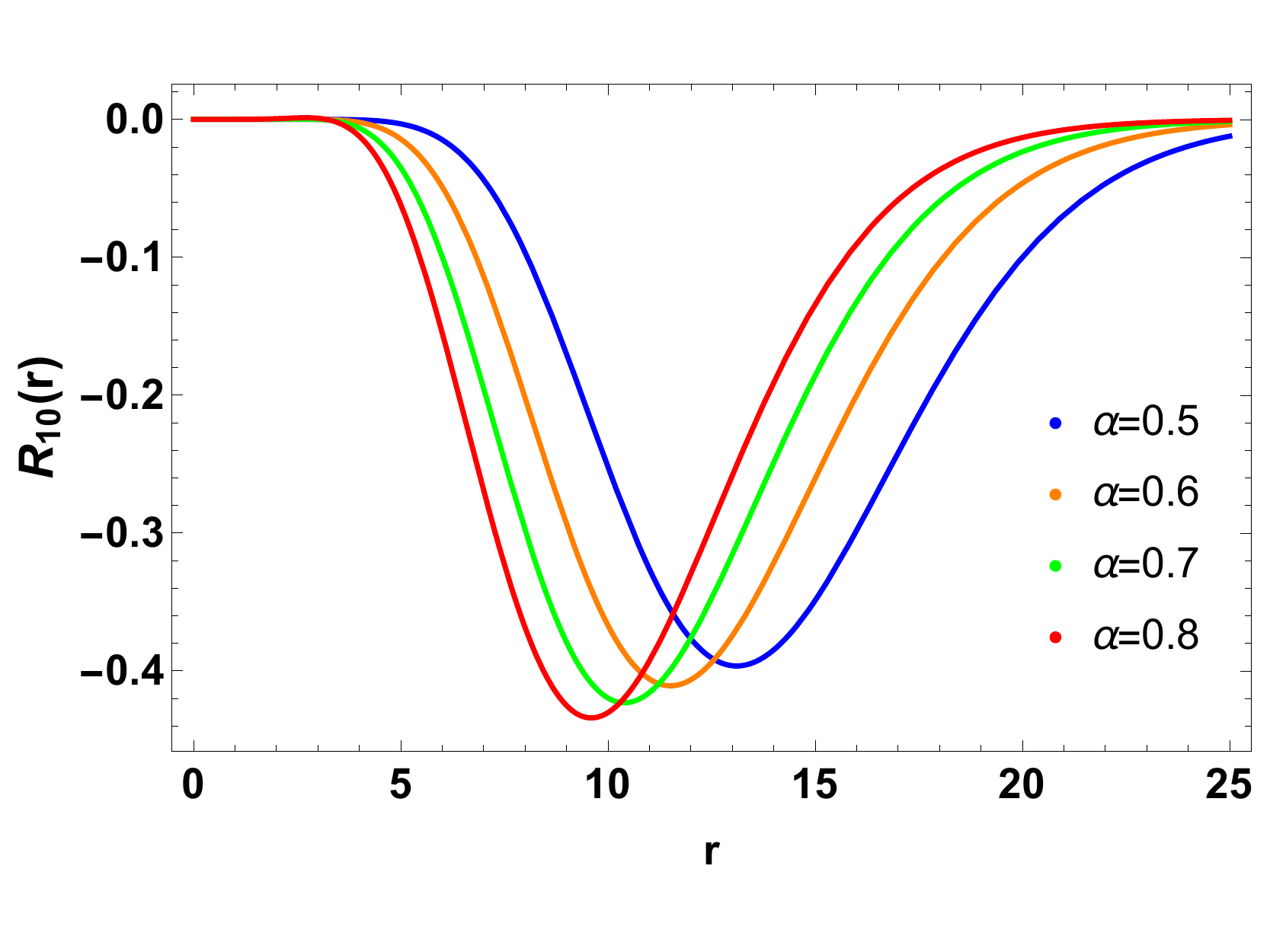}
    \includegraphics[height=4.5cm,width=5cm]{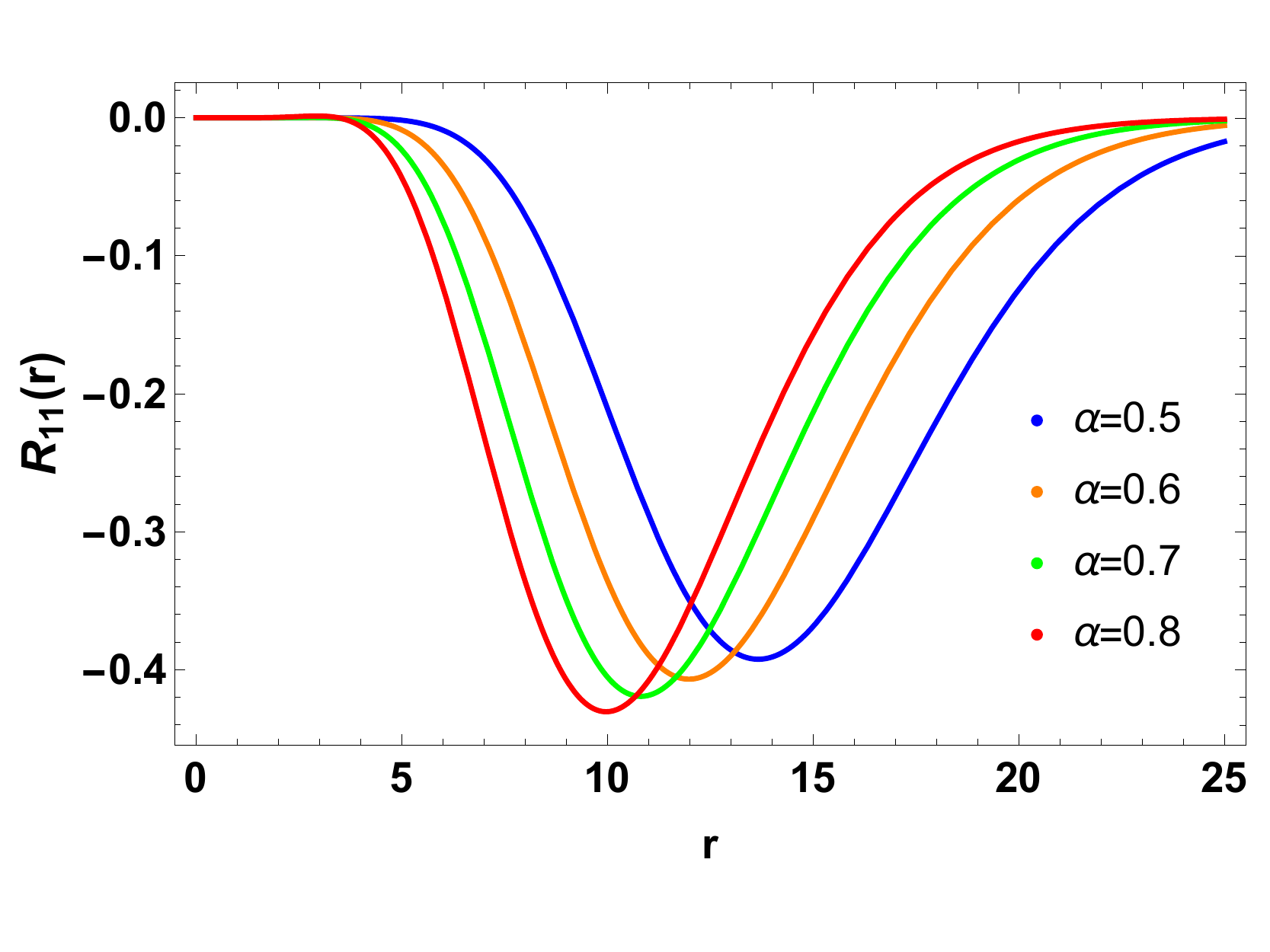}\\ \vspace{-0.3cm}
    (a) \hspace{4.5cm} (b) \hspace{4.5cm} (c)
    \end{center}\vspace{-0.5cm}
    \caption{Radial wave eigenfunctions of heavy mesons (Bottomonium) when  $m_{b}=4.823\,\text{Gev}, \; \mu= 2.4115\,\text{Gev}, \; W_{1}=0.2\,\text{Gev}, \; W_{2}=1.569\,\text{Gev}, \;\delta=0.378\,\text{Gev},\; \text{and}\; \hbar=1$. Fig. (a) corresponds to the state $n=0$ and $l=0$. On the other hand, Fig (b) is the state $n=1$ and $l=0$. The state $n=1$ and $l=1$ is in Fig. (c).}
    \label{fig2}
\end{figure*}

Comparing to Eq. (\ref{eq9}) with Eq. (\ref{NUEq}), it follows that
\begin{align}
    \tilde{\tau}(x)=2x,\quad \sigma(x)= x^2 \quad \text{and}  \quad \tilde{\sigma}(x)=-\tilde{\epsilon}+\tilde{\beta}_{1}x-\tilde{\beta}_{2}x^2.
\end{align}
That explicitly shows that Eq. (\ref{NUEq}) satisfies the requirement of the NU approach. Furthermore, it is also worth noting that $\tilde{\sigma}(x)$ and $\sigma(x)$ are polynomials of at most second degree, and $\tilde{\tau}(x)$ is at most polynomials of the first degree. The NU method is popular among mathematical scientists and related disciplines. Several authors have used this method to solve problems of similar interest \cite{karayer,Yasuk,Onate}. Although the method is quite popular, it will be useful to highlight some details to make our article self-contained. For this reason, we will detail this approach in the appendix (please, verify the appendix). Following the steps described in the appendix [Eqs. (\ref{eqA1}-\ref{eqA7})], one obtains that the self-energies and eigenfunctions of our system are, respectively,
\begin{equation} \label{eq10}
E_{n l}=\frac{3 W_{1}}{\delta}-\frac{\alpha^{2}\hbar^{2}}{8 \mu} \left[ \frac{\frac{6 \mu W_{1}}{\alpha^{2} \delta^{2} \hbar^{2}}+\frac{2 W_{2}}{\alpha^{2}\hbar^2}}{n+\frac{1}{2}+\sqrt{\frac{1}{4}+\frac{2 \mu W_{1}}{\delta^{3} \alpha^{2} \hbar^{2}}+\frac{l(l+1)}{\alpha^2}}}\right]^2
\end{equation}
and 
\begin{equation}\label{REq0}
R_{n l}(r)=N_{n l}r^{\frac{\tilde{\beta}_{1}}{2 \sqrt{\tilde{\epsilon}}}} \text{e}^{-r}L_{n}^{\frac{\tilde{\beta}_{1}}{ \sqrt{\tilde{\epsilon}}}}(2\sqrt{\tilde{\epsilon}}r).
\end{equation}
Figs. \ref{fig1} and \ref{fig2} expose the behavior of the radial wave eigenfunctions for the Charmonium and Bottomonium states of heavy mesons.

\section{The quantum information entropy}

Since the seminal paper by Claude E. Shannon on the mathematical theory of communication \cite{Shannon}, there has been a growing interest in the studies of information entropies, e. g., see Refs. \cite{FCLima1,COEdet4,Edet2,Lima2,Lima3}. In quantum mechanics, part of this interest is because the quantum information entropies are related to the uncertainty measures of the quantum system \cite{Dong1,Dong2,Dong3}. Furthermore, this entropy tells us how good the description of the quantum states of the theory is \cite{FCLima1,COEdet4,Lima2,Lima3}.

Quantum information entropies are related to uncertainty measures through the well-known Bialynicki-Birula and Mycielski (BBM) relation \cite{BBM}, i. e.,
\begin{align}\label{BBMrelation}
    S_{\textbf{r}}+S_{\textbf{p}}\geq\, D\, (1+\text{ln}\pi).
\end{align}
 Here, $S_{\textbf{r}}$ is the information entropy at position space; $S_{\textbf{p}}$ is the information entropy at momentum space (or reciprocal space). Besides, $D$ describes the spatial dimension of the system, e. g., $D=1$ for the one-dimensional case, $D=2$ for two-dimensional one, and $D=3$ for three-dimensional one. In particular, the BBM relation to the quantum theory of the heavy mesons in the presence of point-like defects is $S_{\textbf{r}}+S_{\textbf{p}}\geq\, 6.43419$. It is interesting to mention that the Bialynicki-Birula and Mycielski relation (\ref{BBMrelation}) is a sophisticated version of Heisenberg's uncertainty principle. Thus, the results of quantum information entropy must respect this condition.

For the quantum system discussed earlier (see Sec. \ref{II}), the information entropy is
\begin{align}\label{Sr}
    S_\textbf{r}=-\int_{0}^{2\pi}\int_{0}^{\pi}\int_{0}^{\infty}\vert\psi(\textbf{r},t)\vert^2\, \text{ln} [\vert\psi(\textbf{r},t)\vert^2] r^2\sin\theta\,dr\,d\theta\,d\phi.
\end{align}
Meanwhile, the information entropy at the reciprocal space is
\begin{align}\label{Sp}
    S_\textbf{k}=-\int_{0}^{2\pi}\int_{0}^{\pi}\int_{0}^{\infty}\, \vert\psi(\textbf{k},t)\vert^2\, \text{ln} [\vert\psi(\textbf{k},t)\vert]^2\, k^2\sin\theta_k\,dk_r\,dk_\theta\,dk_\phi&.
\end{align}

The spherical wave eigenfunction at reciprocal space is
\begin{align}\label{transf}
    \psi(\textbf{k},t)=\frac{(2\pi)^{3/2}}{k^{-1/2}}\sum_{l=0}^{\infty}(-i)^l\sum_{m=-l}^{l}Y_{lm}(k_\theta,k_\phi)\int_{0}^{\infty}J_{\frac{1}{2}+l}(k_r r)r^{3/2}\,dr\int_{0}^{2\pi}\int_{0}^{\pi}\, \psi(\textbf{r},t)Y_{lm}(\theta,\phi)r^2\sin\theta\, d\theta\,d\phi.
\end{align}
Eq. (\ref{transf}) is Hankel's transform of the wave eigenfunction $\psi(\textbf{r}, t)$. Hankel's transform is a transform whose kernels are Bessel functions. Thus, Hankel's transform also is called Bessel's transform. In general, Hankel's transform is the two-dimensional Fourier transform of a circularly symmetric function or also spherical three-dimensional functions \cite{Arfken}.

Considering the definitions presented in Eqs. (\ref{Sr}-\ref{Sp}), we can study the quantum information entropy of heavy mesons. Thus, we begin our study by discussing the quantum information entropy of the Charmonium state.

\subsection{Quantum information entropy for the Charmonium}

Quarkonia, in particle physics, are hadronic states made of a heavy quark-antiquark pair. For the vector case, they are the charmonium meson $J/\psi$ (made up of a $\overline{c}c$ pair) and bottomonium meson $\Upsilon$ (made up of a $\overline{b}b$ pair) and their corresponding radial excitations. In fact, in the literature, the word quarkonium refers to the state charmonium and bottomonium \cite{Roldao}. Briefly, Charmonium is a bound state of a quark and charmed antiquark. We found, in the literature, several investigations discussing the Charmonium states. For example, some investigations study about properties of hot gluonic plasma \cite{Datta}, the classification of states of mesons and their electromagnetic decays \cite{Novikov}, and the collision of heavy ions \cite{Rapp}.

Motivated by these applications, let us now study the quantum information of this state. Thus, we will now particularize our study to the quantum information entropies of the Charmonium state. To inspect Charmonium's information it is necessary to assume $m_{c}=1.209\,\text{Gev}, \; \mu= 0.6045\,\text{Gev}, \; W_{1}=0.20\,\text{Gev}, \; W_{2}=1.244\,\text{Gev}, \;\delta=0.231\,\text{Gev}, \text{and}\; \hbar=1$. Using these values, we describe eigenfunctions (\ref{rpsi}) of the Charmonium eigenstates. With the wave eigenfunctions (\ref{rpsi}), one constructs the probability densities of the theory, i. e., $\vert\psi(\textbf{r}, t)\vert^2$. Substituting the probability density in the definition of information entropy (\ref{Sr}) comes to the numerical results of the quantum information entropy at position space. In Tab \ref{tabI}, we expose the numerical result of the information at the position space. To calculate the information entropy associated with the momentum of the Charmonium, the Hankel transform (\ref{transf}) of the eigenfunctions (\ref{rpsi}) is calculated. Substituting the spherical wave eigenfunctions (\ref{transf}) in the definition (\ref{Sp}), one obtains the numerical results of quantum information entropy at the momentum space shown in Tab. \ref{tabI}.

\begin{table*}[!ht]
    \centering
    \caption{Quantum entropy information for the heavy mesons (Charmonium state)}
    \resizebox{11cm}{5.5cm}{%
    \begin{tabular}{|c|c|c|c|c|c|}\hline\hline
        Eigenstates $(n,l,m)$ & Parameter $\alpha$ & $S_\textbf{r}$ & $S_\textbf{p}$ & $S_\textbf{r}+S_\textbf{p}$ & BBM relation \\ \hline\hline
      \multirow{4}{*}{(0,0,0)} & 0.5 & 9.35657 & -0.01643 & 9.34014 & 6.43419 \\ \cline{2-6}
         & 0.6 & 8.67692 & -0.00799 & 8.66893 & 6.43419 \\ \cline{2-6}
         & 0.7 & 8.11763 & 0.01172 & 8.12935 & 6.43419 \\ \cline{2-6}
         & 0.8 & 7.64474 & 0.04550 & 7.69024 & 6.43419 \\ \hline\hline
         \multirow{4}{*}{(1,0,0)}   & 0.5 & 8.61339 & -1.65703 & 6.95636 & 6.43419 \\ \cline{2-6}
         & 0.6 & 8.11752 & 0.09796 & 8.21548 & 6.43419 \\ \cline{2-6}
         & 0.7 & 7.72505 & 0.13561 & 7.86066 & 6.43419 \\ \cline{2-6}
         & 0.8 & 7.40575 & 0.17689 & 7.58264 & 6.43419 \\ \hline\hline
         \multirow{4}{*}{(1,1,0)}   & 0.5 & 8.18144 & -0.99643 & 7.18501 & 6.43419 \\ \cline{2-6}
         & 0.6 & 7.68558 & -0.90431 & 6.78127 & 6.43419 \\ \cline{2-6}
         & 0.7 & 7.29310 & -0.83069 & 6.46241 & 6.43419 \\ \cline{2-6}
         & 0.8 & 7.16383 & -0.72070 & 6.44313 & 6.43419 \\ \hline\hline
         \multirow{4}{*}{(1,1,1)} & 0.5 & 8.20998 & -1.47165 & 6.73833 & 6.43419 \\ \cline{2-6}
         & 0.6 & 7.78339 & -1.05141 & 6.73198 & 6.43419 \\ \cline{2-6}
         & 0.7 & 7.43900 & -1.00400 & 6.43500 & 6.43419 \\ \cline{2-6}
         & 0.8 & 7.15430 & -0.69458 & 6.45972  & 6.43419 \\ \hline
    \end{tabular}}
    \label{tabI}
\end{table*}

The numerical results in Tab. \ref{tabI} suggest that the quantum information entropy decreases at position space as the topological defect increase. In other words, when the point-like global monopole defect increase, the greater the quantum information of the Charmonium states. As a consequence of this behavior, the uncertainty measures related to the position measurements of the heavy mesons will be smaller when the $\alpha$ parameter increases. In contrast, as the point-like global monopole defect becomes large, the information entropy associated with the momentum of the mesons decreases. Thus, one can conclude that when the PGM defect increase, the momentum uncertainty measures of the particles in the Charmonium eigenstate decrease. Furthermore, we note that the BBM relation to the Charmonium state of the quarks remains valid for all eigenstates. Finally, it is possible to note that for higher energy levels, when the parameter $\alpha\to 1$, BBM's relation tends to the minimum uncertainty relation, i. e., $S_\textbf{r}+S_\textbf{p}\approx 6.43419$.

\subsection{Comments on Bottomonium Quantum Information Entropy}

For the Charmonium state $m_{c}=1.209\,\text{Gev}, \; \mu= 0.6045\,\text{Gev}, \; W_{1}=0.20\,\text{Gev}, \; W_{2}=1.244\,\text{Gev}, \;\delta=0.231\,\text{Gev}, \text{and}\; \hbar=1$. For the Bottomonium status, these parameters assume the following values: $m_{b}=4.823\,\text{Gev}, \; \mu= 2.4115\,\text{Gev}, \; W_{1}=0.2\,\text{Gev}, \; W_{2}=1.569\,\text{Gev}, \;\delta=0.378\,\text{Gev},\; \text{and}\; \hbar=1$. Notice that from the Charmonium state to the Bottomonium state, there is an increment of these values and the wave function profile is invariant. Thus, by inspection, one perceives that the quantum information entropy of the Bottomonium state has a similar behavior to the Charmonium state. Indeed, the quantum information entropy $\text{S}_r$ decreases at the position space as the parameter $\alpha$ (the PGM defect) increases. Meanwhile, the information entropy $S_\textbf{p}$ increases at the momentum space when the parameter $\alpha$ decreases. The variation of the $S_\textbf{r}$ and $S_\textbf{p}$ information must vary so that the BBM relation remains valid. These results of the quantum information entropy of the Bottomonium state (and Charmonium presented previously) of the quarks agree with the predictions presented in Ref. \cite{Roldao} using the configurational entropy formalism.

\section{Conclusion}

 In this work, we investigated the Charmonium and Bottomonium states that describe heavy mesons using Schr\"{o}dinger's theory in the presence of a PGM defect. To carry out this study, we considered a curved background coupled with the PGM defect. Using the Nikiforov-Uvarov (NU) formalism, the quantum eigenstates of the heavy mesons were studied. Then, considering the wave functions that describe the quarkonium eigenstates, we investigate the QIE of heavy mesons using Shannon's formalism.

Analyzing quantum self-states that describe the heavy mesons, one notes that the radial eigenfunctions are the associated Laguerre polynomials. Furthermore, to investigate the self-states, the NU method is considered. This method leads us to the energy spectrum shown in Eq. (\ref{eq10}). This spectrum tells us that in the classical limit $E_{n,l\to\infty}\approx 3W_1/\delta$, i. e., the energy will depend purely on $W_1$, the contribution of short-range of the potential. Meanwhile, the fundamental states will depend on short- and long-range contributions of Cornell's potential. Besides, one perceives that the PGM defect alters the quantum eigenstates of the particles as it varies. Indeed, when $\alpha$ increases, the wave function amplitude increases. Thus, the larger the $\alpha$ parameter, the greater the probability of finding quarkonium in regions close to the PGM defect.

Analyzing the QIE, we perceive that the quantum information decreases (at the position space) as the PGM defect increases. On the other hand, it increases if the contribution of the PGM defect decreases. In contrast, at the reciprocal space, an opposite behavior is observed. Furthermore, it is necessary to mention that behaviors of the quantum information are consequences of the wave function profiles that describe the Charmonium and Bottomonium states. As a direct consequence, also, of this result, it is noted that the momentum's uncertainty measures of the heavy mesons decrease as the position uncertainties increase. Thus, it is possible to notice that the BBM uncertainty relation remains valid for all energy levels. Furthermore, the BBM relation tends to a minimum value for higher energy states when $\alpha\to 1$. This occurs because, in this scenario, the PGM defect traps the heavy meson.

This work has some prospects. Among them, a direct perspective of this work is the study of the quantum dynamics of heavy mesons in the presence of other classes of defects, e. g., string-like or domain wall defects. Another perspective is the study of the thermodynamic and statistical properties of these particles when subjected to thermal baths. We hope to carry out these studies soon.


\section*{Authors Declaration}
\subsection*{Funding}
This research has been carried out under LRGS Grant LRGS/1/2020/UM/01/5/2 (9012-00009) Fault-tolerant Photonic Quantum States for Quantum Key Distribution provided by Ministry of Higher Education of Malaysia (MOHE).

\subsection*{Conflicts of interest/Competing interest}
All the authors declared that there is no conflict of interest in this manuscript.

\subsection*{Acknowledgements}

C. O. Edet acknowledges eJDS (ICTP). C. A. S. Almeida thanks to Conselho Nacional de Desenvolvimento Cient\'{\i}fico e Tecnol\'{o}gico (CNPq), n$\textsuperscript{\underline{\scriptsize o}}$ 309553/2021-0. F. C. E. Lima is grateful to Coordena\c{c}\~{a}o de Aperfei\c{c}oamento de Pessoal de N\'{i}vel Superior (CAPES), n$\textsuperscript{\underline{\scriptsize o}}$ 88887.372425/2019-00.

\appendix

\section{Review of the Nikiforov-Uvarov (NU) Method}

Initially proposed by Nikiforov and Uvarov, the NU method seeks to solve Hypergeometric-type differential equations of the form of Eq. (\ref{eq9}) \cite{NU1,NU2}. One can obtain the solutions of Eq. (\ref{eq9}) employing the trial wave function:
\begin{align} \label{eqA1}
P(x)=\Phi(x) y_{n}(x),
\end{align}
which reduces the differential equation (\ref{eq9}) to a Hypergeometric-type differential equation of the form:
\begin{align} \label{eqA2}
\sigma(x) y_{n}^{\prime \prime}(x)+\tau(x) y_{n}^{\prime}(x)+\lambda y_{n}(x)=0.
\end{align}

Allow us to highlight that the function $\Phi (x)$ is defined as
\begin{align} \label{eqA3}
\frac{\Phi^{\prime} (x)}{\Phi (x)}=\frac{\pi (x)}{\sigma(x)},
\end{align}
where $\pi(x)$ is a polynomial of first-degree. Meanwhile, the second term in Eq. (\ref{eqA1}) is the hypergeometric function with its polynomial solution given by Rodriques relation, i. e., 
\begin{align}\label{eqA4}
y_{n}(x)=\frac{B_{n}}{\rho(x)}\dfrac{d^n}{ds^n}\left[\sigma^{n}\rho(x)\right]
\end{align}
The term $B_n$ is the normalization constant, and $\rho (x)$ is known as the weight function, which in principle must satisfy the condition given; 
\begin{align}\label{eqA5}
\dfrac{d}{ds}\left[\sigma\rho(x)\right]=\tau(x)\rho(x)
\end{align}
where $\tau(x)=\tilde{\tau}(x)+2 \pi (x)$.

Naturally, we note here that the derivative of $\tau(x)$ should be $\tau(x) < 0$. Thus, one can obtain the eigenfunctions and eigenvalues using the expression defined by $\pi(x)$  and $\lambda$, i. e.,
\begin{align}\label{eqA6}
\pi (x) =&\frac{\sigma^{\prime}(x)-\tilde{\tau}(x)}{2} \pm \sqrt{\left(\frac{\sigma^{\prime}(x)-\tilde{\tau}(x)}{2}\right)^{2}-\tilde{\sigma}(x)+k \sigma(x)}
\end{align}
and
\begin{align}\label{eqA6.1}
\lambda=k+\pi^{\prime}(x).
\end{align}

Considering the discriminant of the square root [in Eq. (\ref{eqA6})] equal to zero, we obtain the value of $k$. In this case,  the new eigenvalue equation is
\begin{align}\label{eqA7}
\lambda + n \tau^{\prime}(x) + \frac{n(n-1)}{2} \sigma^{\prime \prime} (x) = 0, \hspace{0.5cm} \text{with} \hspace{0.5cm} n=0,1,2,\dots
\end{align}
 
\section{Solutions in Detail}

Substituting $\tilde{\sigma}(x)=-\tilde{\epsilon}+\tilde{\beta}_{1}x-\tilde{\beta}_{2}x^2$ into Eq. (\ref{eqA6}), we arrived at
\begin{align} \label{eqB1}
\pi (x) = \pm \sqrt{\tilde{\epsilon}-\tilde{\beta}_{1}x+(\tilde{\beta}_{2}+k) x^{2}}.
\end{align}

The discriminant of the quadratic expression under the square root above is
\begin{align}\label{kEq}
    k=\frac{\tilde{\beta}_{1}^{2}-4 \tilde{\beta}_{2} \tilde{\epsilon} }{4\tilde{\epsilon}}.
\end{align}

Substituting Eq. (\ref{kEq}) in (\ref{eqB1}), one obtains
\begin{align}
    \pi (x) = \pm \left(\frac{\tilde{\beta}_{1}x}{2\sqrt{\tilde{\epsilon}}}-\frac{\tilde{\epsilon}}{\sqrt{\tilde{\epsilon}}}\right),
\end{align}
where
\begin{align}
    \pi^{\prime} (x) = -\frac{\tilde{\beta}_{1}}{2\sqrt{\tilde{\epsilon}}}.
\end{align}
Thus, we can conclude that
\begin{align}\label{eqB2}
\tau(x)=2x-\frac{\tilde{\beta}_{1}x}{\sqrt{\tilde{\epsilon}}}+\frac{\tilde{\epsilon}}{\sqrt{\tilde{\epsilon}}} \quad \text{and}  \quad \tau^{\prime}(x) = 2-\frac{\tilde{\beta}_{1}}{\sqrt{\tilde{\epsilon}}},
\end{align}
which leads us to
\begin{align}\label{eqB3}
\frac{\tilde{\beta}_{1}^{2} - 4 \tilde{\beta}_{2} \tilde{\epsilon}}{4 \tilde{\epsilon}}-\frac{\tilde{\beta}_{1}}{2\sqrt{\tilde{\epsilon}}}=\frac{n \tilde{\beta}_{1}}{2 \sqrt{\tilde{\epsilon}}}-n^2-n
\end{align}
Eq. (\ref{eqB3}) yields the energy equation of the Cornell potential presented in Eq. (\ref{eq10})


\end{document}